\documentclass[a4paper]{aa}
\usepackage{ffbib, epsfig, times}

\newcommand{\cq}{\mbox{$\chi^2$}}
\begin{document}
\thesaurus{07(08.09.2 AD Leo; 08.12.1; 08.01.2; 08.03.5; 13.25.5)}

\title{The corona of the dMe flare star AD~Leo}

\author{F.~Favata\inst{1} \and G.~Micela\inst{2} \and
  F.~Reale\inst{3}}

\institute{Astrophysics Division -- Space Science Department of ESA, ESTEC,
  Postbus 299, NL-2200 AG Noordwijk, The Netherlands
\and
Osservatorio Astronomico di Palermo, 
Piazza del Parlamento 1, I-90134 Palermo, Italy 
\and
Dip.\ Scienze FF.\ \& AA., Sez. Astronomia, Univ.\ Palermo,
Piazza del Parlamento 1, I-90134 Palermo, Italy
}

\offprints{F. Favata} \mail{ffavata@astro.estec.esa.nl}

\date{Received 24 August 1999; Accepted 29 November 1999}

\maketitle 
\begin{abstract}

  We have systematically studied the X-ray emission (both the
  quiescent component and the flares) of the dM3e star AD~Leo,
  analyzing the existing observations from the \emph{Einstein}~IPC,
  ROSAT PSPC and ASCA SIS instruments. Using a consistent method which
  explicitly considers the possibility of sustained heating we have
  analyzed the six flares which have sufficient statistics, deriving
  constraints on the physical parameters of the flaring regions.  In
  all cases the flaring loops are likely compact ($L \simeq 0.3~R_*$),
  and confined to a rather narrow range of sizes, incompatible with
  the large ($L \ga R_*$) tenuous loops claimed by previous analyses
  of flares on AD~Leo and other similar stars. The flaring loops
  appear to have a larger cross section ($\beta = r/L \simeq 0.3$)
  than customarily assumed (e.g.\ $\beta \simeq 0.1$).  All flares
  show evidence of significant heating during the decay phase.
  Although the derived peak pressures are high (up to $P \simeq
  10^4$~dyne~cm$^{-2}$) with a peak temperature of $\simeq 50$~MK, the
  magnetic fields required to confine such loops and to produce the
  observed flare luminosity are relatively modest ($B \simeq
  1$--$2$~kG) and fully compatible with the photospheric magnetic
  fields measured in several flare stars.  If the narrow range of loop
  sizes obtained is extrapolated to the quiescent structures
  responsible for the active corona, the latter can be naturally
  scaled up from the solar case through a modest (a factor of 10)
  increase in pressure in otherwise solar-like active structures with
  a small surface filling factor ($\simeq 5$\%). The quiescent
  component of the corona shows no evidence for abundance
  peculiarities with respect to the photosphere, and the quiescent
  coronal luminosity is remarkably constant (with variations of less
  than a factor of 2) across the almost 20~yr span of the observations
  discussed here.


\end{abstract}

\section{Introduction}
\label{sec:intro}

While the solar analogy is almost universally accepted as the starting
point for the modeling of stellar coronae, its scaling to the much
higher activity levels seen in active stars is still a debated
question.  The Sun displays, even at the maximum of its 11~year cycle,
a rather low level of activity, when compared to the most active
stars; dMe dwarfs (flare stars) on the other hand exhibit quiescent
activity levels which are up to three orders of magnitude higher (n
terms of the surface flux) than the Sun at minimum, and still a factor
of hundred higher when compared with the Sun at maximum.

If the solar corona is considered in an integrated fashion (as one has
to do for unresolved stellar sources) different areas of the Sun will
dominate the spectrum depending on the band of interest.
\cite*{opr+99} have for example examined the Sun in an intermediate
activity state, showing that active regions (defined as regions whose
X-ray surface luminosity in the Yohkoh/SXT thin aluminum filter is at
least 1\% of the peak surface luminosity of the Sun at the given time)
occupy only 2--3\% of the solar surface, with the rest of the solar
surface covered with low surface-brightness structures (quiet
regions).  Given that the active regions have significantly higher
temperatures than the quiet regions, they dominate the spectrum at
different energies: notwithstanding their very low surface brightness,
quiet regions still dominate the integrated spectrum below $\simeq
0.5$~keV, while the few (seven in the case examined by \cite{opr+99})
active regions dominate the spectrum at energy $\ga 0.5$~keV, where
the quiet regions are essentially invisible (\cite{por+99}). The
active regions dominating the emission above $E \simeq 0.5$~keV have
characteristic sizes (loop semi-lengths) $L \la 0.2~R_\odot$ (e.g.\ 
\cite{mew92} and reference therein).

Several options are in principle available for scaling the solar
corona up to the much higher X-ray luminosity seen in very active
stars. One possibility is to augment the surface filling factor of the
active region structures, up to complete coverage of the solar
surface, without significantly changing their characteristics (size
and pressure). This would lead to a maximum possible increase in the
X-ray luminosity of a factor of approximately 100, thus naturally
explaining the X-ray luminosity ($\simeq 5 \times
10^{28}$~erg~s$^{-1}$) of intermediate-activity solar-type stars (as
discussed in detail by \cite{dpo+99}).  To explain the higher X-ray
luminosity (up to few times $10^{30}$~erg~s$^{-1}$) observed in the
most active solar-type stars different mechanisms must be at work. The
coronal structures must either be significantly larger than in the
solar case, and/or the plasma pressure must increase. Large coronal
structures ($L \ga R_*$) are implied by the results of several studies
of the decay phase of stellar flares. If the active regions of X-ray
luminous stars are so large, the increase in available volume,
together with a high volume filling factor, could indeed lead to the
higher X-ray luminosities observed in the most active stars without
significantly increasing the plasma pressure -- and the observed
saturation could be explained in terms of a maximum coronal filling
factor. At the same time, however, increasing the coronal pressure $n$
is -- given that the X-ray emissivity of the plasma at typical coronal
temperatures scales as $n^2$ -- a highly effective way to increase the
X-ray luminosity.

The question of the scaling of the corona to higher X-ray luminosity
is relevant to stellar evolution in general: higher-pressure
structures imply larger confining magnetic fields, which in turn will
have an influence on the convection region (magnetic fields modify the
heat-transport capacity of the convective envelope), and thus on the
structure of the star as such. High internal magnetic fields may even
prevent a low-mass star from becoming fully convective. The structure
of the magnetic field will also have an influence on the loss of
angular momentum through magnetic braking, and thus again on the
evolution of the internal structure of the star, as well as, likely,
on the depletion pattern of light elements (\cite{vzm+98}).  Finally,
the issue of whether the frequent and intense flares seen in very
active stars are produced within closed coronal structures (as opposed
to the production of intense flares in freely expanding plasmoids,
i.e.\ large-scale coronal mass ejections) is relevant to the question
of the contribution of the plentiful low-mass flare stars to the
chemical evolution of our Galaxy.

Lacking spatial resolution, no direct information about the size of
coronal structures can be obtained for stars (although this is
sometimes possible for eclipsing binaries, see \cite{sch98} and
\cite{sf99}). The upcoming generation of high spectral resolution
X-ray missions (\emph{Chandra} and XMM) will offer the possibility of
direct measurements of density-sensitive spectral diagnostics at
different coronal temperatures, and thus of estimating (using the
differential emission measure determination from the line fluxes) the
volume of the emitting region as a function of temperature. Even with
these future large-area X-ray observatories, however, only a limited
number of bright sources will yield enough photons to produce
sufficiently high $S/N$ spectra. For the most part, thus, information
about the structuring of the corona on the majority of active stars
will still have to rely, for the foreseeable future, on indirect
methods.

The study of the decay phase of stellar flares is up to now the main
available tool for the determination of the characteristic sizes of
coronal structures, although in principle it only gives information
about the flaring regions. While many flares have been studied in
these terms, most studies have been limited to the analysis of
individual flaring events, whose size may well be peculiar and not
representative of the typical scales of the corona as a whole, and in
particular of its quiescent structures.  For the few stars for which a
significant number of flares has been observed, a systematic,
homogeneous study can provide a more general view of the the
characteristic sizes of coronal structures; based on the solar
analogy, where the same active regions which dominate the
high-temperature emission measure are responsible for the flares, the
flaring loops can then be used, qua size, as proxies for the active
corona as a whole.

\begin{figure*}[!thbp]
  \begin{center}
    \leavevmode \epsfig{file=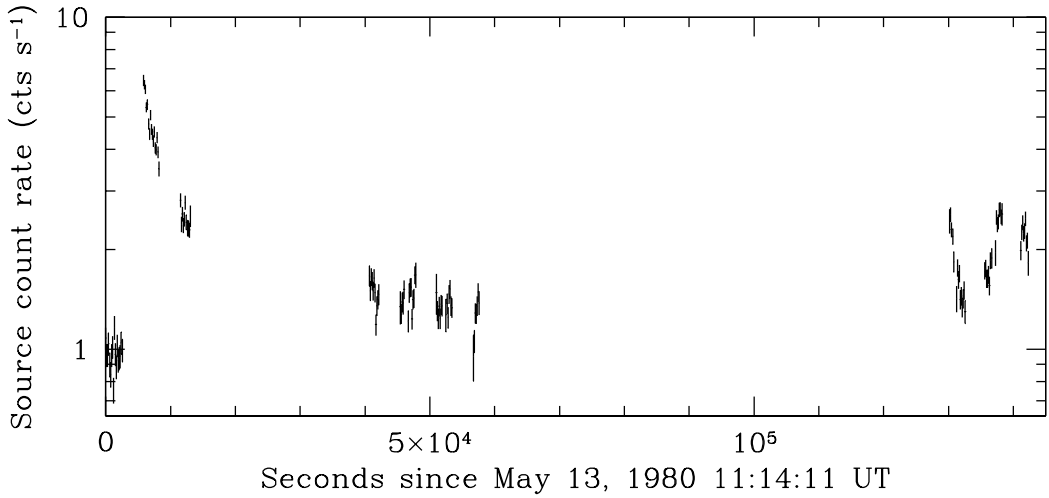, width=15.0cm, bbllx=20pt,
      bblly=400pt, bburx=600pt, bbury=700pt, clip=}
    \caption{The light-curve of AD~Leo in the \emph{Einstein} IPC
      observation, background-subtracted (background count rate
      $\simeq 0.087$ cts s$^{-1}$) and binned in 150~s intervals.}
    \label{fig:ipclc}
  \end{center}
\end{figure*}

In the present paper we undertake such systematic study on the flare
star AD~Leo (a dM3e star whose characteristics are discussed in detail
in \ref{app:targ}), which is an ideal target given its high X-ray flux
and the significant number of flares observed thus far. The flares we
analyze here have been observed by three different instruments
(\emph{Einstein} IPC, ROSAT PSPC and ASCA SIS), with different
bandpasses and energy resolution. The different bandpasses allow to
explore events across a range of peak temperatures; if the
characteristics of the flaring loops thus derived are similar, this is
a strong indication of the corona being dominated by a class of
structures. We have coherently analyzed all the AD~Leo flares with
sufficient statistics; the main result is that all six flares
originate from similar, compact structures.  Assuming that the loops
which are responsible for the flares are also the main constituent of
the active corona, we then constrain its characteristics, in
particular regarding the filling factor and the coronal pressure. The
paper is structured as follows: Sect.~\ref{sec:analy} describes the
observations and their reduction, the analysis of the individual
flares is discussed in Sect.~\ref{sec:tech} (with specific details
about the analysis of the \emph{Einstein} IPC flare in
\ref{app:ipctech}), the results are discussed in Sect.~\ref{sec:disc},
while Sect.~\ref{sec:concl} contains the conclusions.

\section{Observations and data reduction}
\label{sec:analy}

AD~Leo has been observed on several occasions both in the X-ray and UV
(as described in Appendix~\ref{app:prex}). In the present paper we
have analyzed the \emph{Einstein}, ROSAT and ASCA X-ray observations,
in all of which significant flaring events were detected.
Table~\ref{tab:dets} shows a summary comparison of the characteristics
of the different detectors, observations and number of flaring events
studied in each. All of the observations discussed here were retrieved
from the HEASARC archive, and were analyzed using the {\sc ftools}~4.1
software suite (except for the PSPC flaring spectra which were
analyzed using the {\sc pros} package).  Spectra and light curves were
extracted with the {\sc xselect} package and the spectral analysis was
performed using the {\sc xspec}~10.0 package and the {\sc mekal}
plasma emission model (\cite{mkl95}), with (when necessary) an
interstellar absorption components following the \cite*{mm83} model.
To allow comparison of results obtained with different instruments we
will quote all X-ray luminosities both in the instrument's ``own''
band and in a common 0.5--4.5~keV band.

\begin{table}[htbp]
    \caption{The main characteristics of the \emph{Einstein} IPC,
      ROSAT PSPC and ASCA SIS detectors (spectral band and
      resolution), together with the main features of the AD~Leo
      observation performed by each of them (exposure time, elapsed
      time, number of analyzed flaring events).}
  \begin{center}
    \leavevmode
    \begin{tabular}{rcrrrr}
     &  Band  & Resolution  & $t_{\rm exp}$ & $t_{\rm ela}$ & N.\ flares \\
     & keV    & eV at 1~keV & ks & ks & \\\hline
     IPC & 0.16--3.5 & 1000 & 18.4 & 145 & 1\\
     PSPC & 0.1--2.0 & 450 & 22.5 & 60 & 2\\
     SIS & 0.5--10.0 & 100 & 87.0 & 240 & 3\\
    \end{tabular}
    \label{tab:dets}
  \end{center}
\end{table}

\subsection{The \emph{Einstein} IPC observation}
\label{sec:einstein}

AD~Leo was, on May 13, 1980 (starting at 11:14 UT), the target of a
long IPC observation (18.4~ks effective exposure, $\simeq 145$~ks
elapsed time). This observation was briefly discussed by
\cite*{asg87}, who noted the presence of ``a previously unreported
flare'', and showed that no significant time variability is present
before the flare, while after the event significant variability is
seen on time scales of $\ga 100$~s.  The light curve of the
observation is plotted in Fig.~\ref{fig:ipclc}: the large flare near
the beginning of the observation, with an enhancement of a factor of
approximately 7 in count rate, is evident. 

Somewhat curiously, given that this is the most intense event observed
by the IPC on a flare star, it has never been analyzed previously (nor
to our knowledge reported, apart from the brief mention of
\cite{asg87}). \cite*{scs+90}, in their systematic analysis of all
stellar IPC spectra, note that the spectrum cannot be fit with any of
the models they considered -- i.e.\ one- or two-temperature models or
continuous emission measure models -- most likely because of the
presence of the flaring emission.

We have extracted the spectra (both quiescent and flaring) from a
3~arcmin radius circle centered on the source, and analyzed the
``quiescent'' spectrum extracted in the four observation segments at
around 50~ks from the beginning of the observation. The background was
extracted from a ring with internal and external radii of 3.5 and
5~arcmin. We have, in keeping with the photospheric metallicity and
with the ASCA results, frozen the coronal metallicity to
$0.2~Z_\odot$; no absorbing column density is necessary to fit the
spectrum. The resulting quiescent X-ray luminosity is $L_{\rm X} = 4.4
\times 10^{28}$~erg~s$^{-1}$ in the \emph{Einstein} 0.16--3.5~keV band
(and $3.2 \times 10^{28}$~erg~s$^{-1}$ in the 0.5--4.5~keV band) with
coronal temperatures of 0.3 and 8.5~MK.

\subsection{The PSPC observation}

\begin{figure*}[!thbp]
  \begin{center}
    \leavevmode \epsfig{file=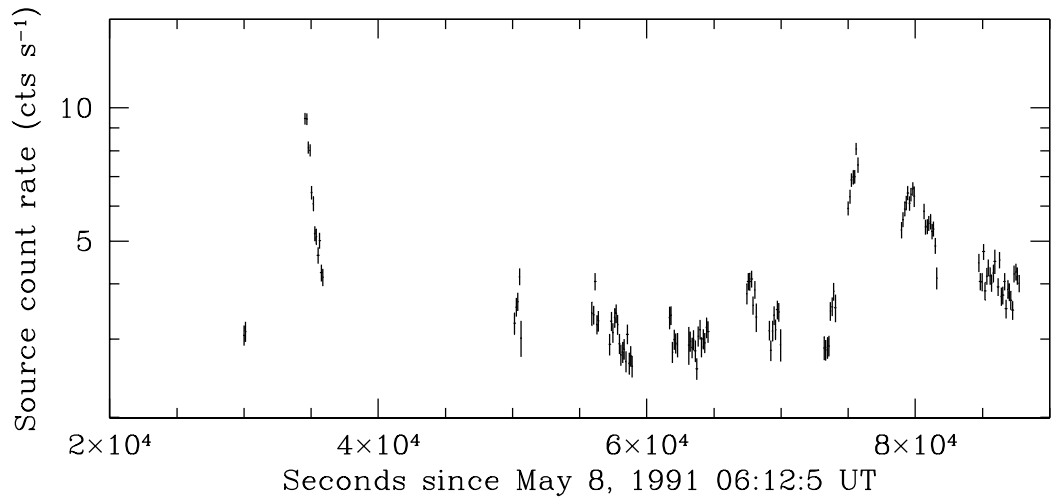, width=15.0cm, bbllx=20pt,
      bblly=400pt, bburx=600pt, bbury=700pt, clip=}
    \caption{The light-curve of AD~Leo in the PSPC
      observation, background-subtracted (background count rate
      $\simeq 0.061$ cts s$^{-1}$) and binned in 120~s intervals.}
    \label{fig:pspclc}
  \end{center}
\end{figure*}

AD~Leo was observed by the ROSAT PSPC in pointed mode on May 8, 1991,
for a total effective exposure of 22.5~ks and a time span of $\simeq
60$~ks. The light-curve (binned in 120~s intervals) is shown in
Fig.~\ref{fig:pspclc}: variability is evident on several time scales,
although a quiescent level of $\simeq 3$ PSPC cts~s$^{-1}$ can be
identified near the middle of the observation. At least two individual
flaring events can be recognized, one starting at $\simeq 28$~ks from
the beginning of the observation and another, longer event starting at
$\simeq 73$~ks, with both rise and decay visible for the second flare.
The average spectrum during the complete observation has been analyzed
both by \cite*{grk+96} and \cite*{smf+99}, as discussed in
Appendix~\ref{app:prex} For the purpose of the flare analysis, source
photons have been extracted from a circular region 3 arcmin in
diameter, and the quiescent spectrum has been obtained from the
segment between 60 and 65~ks, where no flaring activity is evident.

\subsection{The ASCA observation}

\begin{figure*}[!thbp]
  \begin{center}
    \leavevmode \epsfig{file=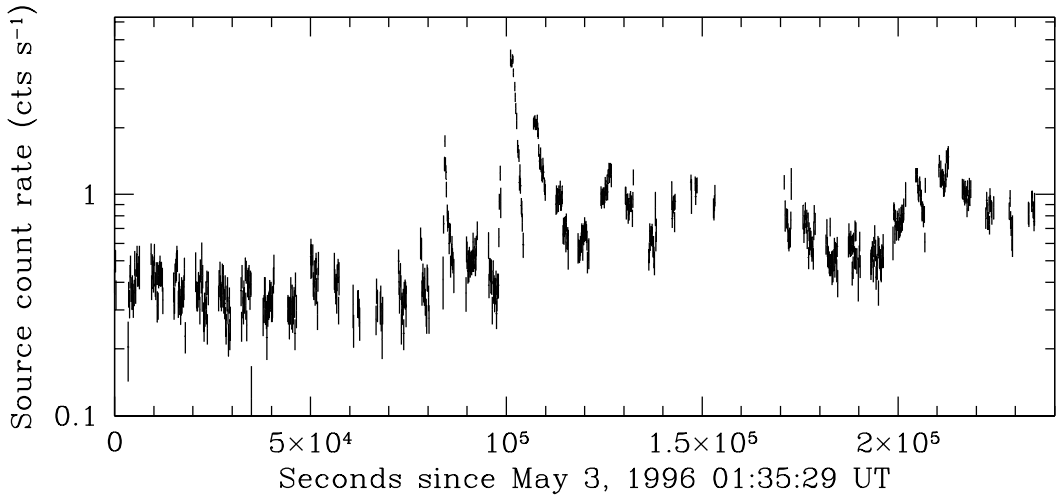, width=15.0cm, bbllx=20pt,
      bblly=400pt, bburx=600pt, bbury=700pt, clip=}
    \caption{The light-curve of AD~Leo for the ASCA observation,
      as seen by the SIS-1 detector, background-subtracted (background
      count rate $\simeq 0.025$ cts s$^{-1}$) and binned in 150~s
      intervals.}
    \label{fig:lctotal}
  \end{center}
\end{figure*}

The ASCA observation of AD~Leo discussed in the present
paper\footnote{ASCA also observed AD~Leo in 1993, in a 2-CCD mode, in
  which the source fell very close to the gap between the two chips,
  making the spectral analysis difficult. The quiescent level during
  was very similar to the 1996 observation, and a small flare is
  present near the end. We will not further consider the 1993
  observation.} was a performed starting on May 3, 1996 01:35 UT for a
total elapsed time of $\simeq 240$~ks, and an on-source time of
$\simeq 87$~ks.  Source photons have been extracted, for both SIS-0
and SIS-1 detectors, from a circular region 3.7~arcmin in radius (36
pixels) centered on the source position, while background photons have
been extracted from the whole CCD chip excluding a circular region
5.7~arcmin in radius, also centered on the source. The SIS-1
background-subtracted light curve for the complete ASCA observation
(binned at 150~s intervals) is shown in Fig.~\ref{fig:lctotal}.

While the first part of the observation is characterized by a
relatively constant light curve, with little variation, significant
variability is present starting at $\simeq 80$~ks from the beginning
of the observations. Several individual flaring events can be
recognized, in particular one starting at $\simeq 80$~ks (flare~1),
one starting at $\simeq 100$~ks (flare~2) and one starting at $\simeq
107$~ks (flare~3). After the flaring activity the light curve is
characterized by a much higher level of variability than before, with
variations in the source count rate of a factor of $\simeq 2$--3 on
time scales of tens of ks evident toward the end of the observation,
i.e.\ comparable to the enhancements observed for the two PSPC flares
we have analyzed. Such flares have too limited statistics to allow
analysis from the ASCA data; the observation shows however that they
are frequent.

\subsubsection{The quiescent emission in the ASCA data}

The first part (up to $\simeq 80$~ks elapsed time) of the ASCA
observation is characterized by a flat light curve, showing little
variability, which allows to study in detail the quiescent coronal
emission from the star. For this purpose, we have extracted spectra
for both the SIS-0 and the SIS-1 detectors for the time interval
0--82~ks from the beginning of the observation, and fit it with a
two-temperature {\sc mekal} spectrum. Given the distance to AD~Leo,
the interstellar column density is expected to be small ($<
10^{19}$~cm$^{-2}$), with no influence on the SIS spectra. Spectra
have been rebinned in energy to variable-size bins containing each at
least 20~cts, and Gehrels statistics have been used in {\sc xspec} to
compute the reduced \cq. The SIS-0 and SIS-1 spectra have been fit
simultaneously to the same source model.

\begin{figure*}[!thbp]
  \begin{center}
    \leavevmode \epsfig{file=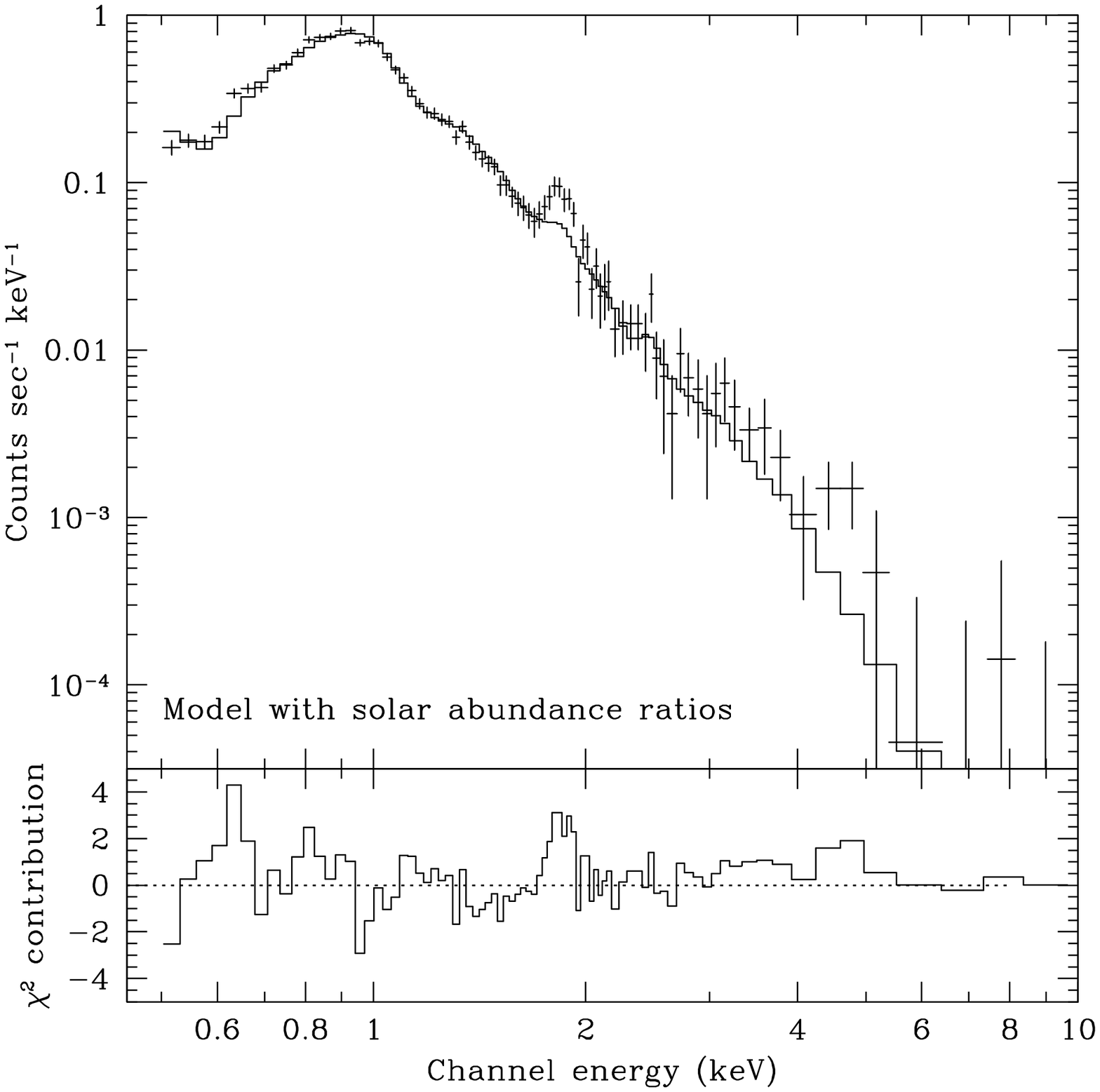, width=8cm, bbllx=15pt,
      bblly=150pt, bburx=600pt, bbury=700pt, clip=}
    \leavevmode \epsfig{file=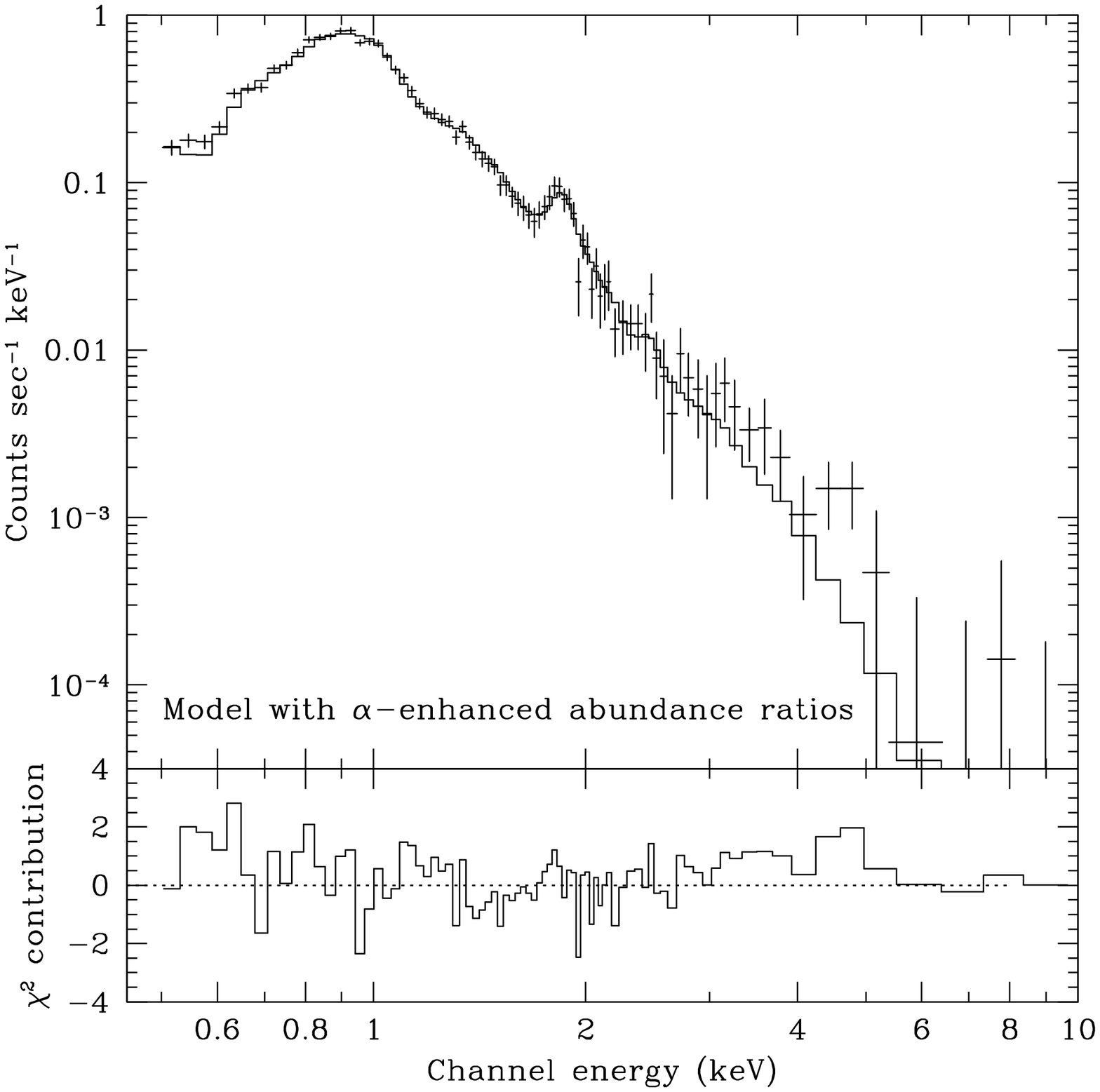, width=8cm, bbllx=15pt,
      bblly=150pt, bburx=600pt, bbury=700pt, clip=}
    \caption{Quiescent SIS spectra of AD~Leo (only the SIS-0 spectrum
      is shown for clarity). The left panel shows the observed
      spectrum together with a best-fit two-temperature {\sc mekal}
      model with all the abundances varying in lockstep (best fit
      ${\rm [Fe/H]} = -0.77$), while the right panel shows a
      two-temperature {\sc mekal} model with the O and Si abundance
      decoupled from the other abundances, with ${\rm [Fe/H]} = -0.68$
      and ${\rm [O/Fe]} \simeq {\rm [Si/Fe]} \simeq 0.4$.}
    \label{fig:qui}
  \end{center}
\end{figure*}

A simple two-temperature fit with abundance ratios fixed to the solar
value converges to ${\rm [Fe/H]} = -0.77$, but fails to yield a
satisfactory fit, with $\cq = 1.51$ over 158 degrees of freedom,
corresponding to a probability level of $\simeq 10^{-5}$. Inspection
of the spectrum (left panel of Fig.~\ref{fig:qui}) shows the presence
of a strong line complex at $E \simeq 1.8$~keV, which is significantly
under-predicted by the model, plus some excess around $E \simeq
0.6$~keV. Given that the line is at the energy of the Si~K complex,
while the low-energy excess is close to the O~K complex energy, we
have fit the data with a two-temperature {\sc mekal} model, with Si
and O abundances decoupled from the other abundances. The resulting
model (shown in the right panel of Fig.~\ref{fig:qui}) does indeed
produce an acceptable fit (reduced $\cq = 0.96$, corresponding to a
probability level of $\simeq 60$\%).  The Si line is now well fit, and
the low-energy excess is reduced. The coronal Fe abundance is ${\rm
  [Fe/H]} = -0.68$, while the both O and Si are $\simeq 2.5$ times
over-abundant with respect to the solar abundance ratios. The fit is
not sensitive to changes in the abundance ratios of the other $\alpha$
elements which might contribute to the X-ray spectrum, i.e.\ Mg, S and
Ca. The best-fit model parameters with relevant confidence ranges are
shown in Table~\ref{tab:qui}. The quiescent X-ray luminosity is
$L_{\rm X} = 5 \times 10^{28}$~erg~s$^{-1}$ in both the instruments
own 0.5--10~keV band and in the 0.5--4.5~keV band.

\begin{table*}[htbp]
    \caption{The spectral parameters derived for the quiescent
      emission of AD~Leo from the analysis of the SIS-0 and SIS-1
      spectra accumulated during the first part of the ASCA
      observation (i.e.\ up to 82~ks elapsed time) using a
      two-temperature {\sc mekal} spectral model. The first row shows
      the results of a fit with abundance ratios fixed to the solar
      values, while the second row shows the best-fit model with O and
      Si abundance decoupled from the rest. The quiescent X-ray
      luminosity corresponding to the spectral parameters shows is $5
      \times 10^{28}$~erg~s$^{-1}$ (both in the 0.5--10~keV and in the
      0.5--4.5~keV bands). Quoted uncertainties correspond to $\Delta
      \cq = 2.7$.}
  \begin{center}
    \leavevmode
    \begin{tabular}{cc|cc|c|c|c|cc}
 $T_1$ & $T_2$ &  $EM_1$ & $EM_2$  &  [Fe/H] &  [O/H] &  [Si/H] &
 $\chi^2$ & DoF\\
\multicolumn{2}{c|}{keV}  & \multicolumn{2}{c|}{$10^{50}$ cm$^{-3}$} &
  &   &   & & \\\hline
$0.37\pm 0.03$ & $0.87 \pm 0.03$ & $28 \pm 1$ & $30 \pm 1$ &
$-0.77 \pm 0.03$ & $={\rm [Fe/H]}$ & $={\rm [Fe/H]}$ & 1.51 & 158 \\
$0.48 \pm 0.08$ & $0.85 \pm 0.05$ & $16 \pm 3$ & $23 \pm 6 $ & $-0.68
\pm 0.08 $ & $-0.27 \pm 0.07$ & $-0.28 \pm 0.08$ & 0.96 & 156 \\
    \end{tabular}
    \label{tab:qui}
  \end{center}
\end{table*}

\section{Flare analysis}
\label{sec:tech}

Various approaches have been developed for the analysis of stellar
flares, in general based on some physical model of the flaring region
and a fit of the observed decay behavior to the model to derive the
``best-fit'' physical parameters. Most of these models assume that the
flare decay is dominated by the characteristics of the flaring region,
with negligible sustained heating during the decay phase.  Among these
are the quasi-static method\footnote{In principle the quasi-static
  formalism, as originally developed, would allow to include sustained
  heating. However, also due to the mathematical complexity of the
  formulation, it has generally been applied assuming free decay of
  the loop.}, first applied by \cite*{om89} for a flare on Algol, and
the method of \cite*{fh90}, applied for example to the analysis of a
flare seen by EUVE on AD~Leo itself (\cite{cfh+97}). This type of
approach has been employed in the literature for the analysis of
several flaring events, as have simpler scaling arguments again based
on the assumption of freely decaying loops; in general, these
analyses, when applied to intense, long-lasting events naturally
result in long ($L \ga R_*$) tenuous loops.

\cite*{rbp+97} have developed an approach to the analysis of the decay
phase of flares based on detailed hydrodynamic modeling of decaying
flaring loops with explicit allowance for sustained heating
(parameterized as an exponential function of time) during the decay
phase.  The method uses as a diagnostic for presence of heating during
the decay phase the slope of the locus of the decay in the $\log
n$--$\log T$ plane (\cite{sss+93}).  An extensive set of hydrodynamic
models of decaying flaring loops has made it possible to derive
empirical relationships between the light curve decay time (in units
of $\tau_{\rm th}$, the loop thermodynamic decay time, \cite{srj+91})
and the slope $\zeta$ of the flare decay in the $\log n$--$\log T$
diagram (using the square root of the emission measure of the flaring
plasma as a proxy to the density).

Application of this approach to the Sun has shown that sustained
heating during the flare decay is common in solar events
(\cite{rbp+97}); when stellar events are analyzed within this
framework, the resulting loop sizes are invariably significantly
smaller ($L < R_*$, i.e.\ ``solar-like'') than when the same events
are analyzed as ``freely decaying''. We have previously applied the
sustained heating framework to the analysis of flaring events on PSPC
flares observed on the dMe stars CN~Leo and AD~Leo itself
(\cite{rm98}), on a large flare observed by SAX on the active binary
system Algol (\cite{fs99}) and on an exceptional flare observed on the
dMe dwarf EV~Lac by ASCA (\cite{frm+99}), with the aim of determining
the characteristic sizes of coronal structures on different types of
active stars. In all cases, these turned out to be compact ($L \la
0.5~R_*$), with evidence for significant sustained heating, so that
the much longer loops ($L > R_*$) implied by the analyses based on the
free decay approach appear to be unrealistic.

We use the \cite*{rbp+97} approach to analyze all the flares discussed
here. The empirical relationship between $\zeta$ and $\tau_{\rm
  LC}/\tau_{\rm th}$ needs to be derived separately for each X-ray
detector, as it will depend on the spectral response. The loop length
is a function of the maximum observed temperature during the flare
$T_{\rm max}$ and of the intrinsic thermodynamic decay time $\tau_{\rm
  LC}$ determined from the observed decay time $\tau_{\rm th}$ and
from $\zeta$, as shown in detail for the ASCA SIS detector in
Sect.~\ref{sec:sisflares}. For instruments with limited spectral
resolution (such as the ROSAT PSPC and the \emph{Einstein} IPC)
\cite*{rm98} have developed an approach based on a principal component
decomposition of the spectrum, which allows optimal use of
low-resolution data.

\subsection{Analysis of the ASCA flares}
\label{sec:sisflares}

To analyze the flares seen with the ASCA-SIS detector we have first
recalibrated the method of \cite*{rbp+97} for use with the ASCA SIS
detectors, and in particular the empirical relationship between the
observed and intrinsic decay time as a function of the slope $\zeta$
in the $\log n$--$\log T$ diagram.

The intrinsic decay time $\tau_{\rm th}$ of a closed coronal loop with
semi-length $L$, and maximum temperature $T_{\rm max}$ is given by
\cite*{srj+91} as:

\begin{equation}
  \label{eq:taulc}
  \tau_{\rm th} = {\alpha L \over \sqrt{T_{\rm max}}}
\end{equation}
where $\alpha = 3.7 ~ 10^{-4} \rm ~ cm^{-1} s^{-1} K^{1/2}$. By means
of a grid of hydrostatic loop models (see \cite{rm98}) we have found
an empirical relationship which links the loop maximum temperature
$T_{\rm max}$, typically found at the loop apex (e.g.\ \cite{rtv78})
to the maximum temperature $T_{\rm obs}$ determined from the SIS
spectrum :

\begin{equation}
  \label{eq:tobs}
  T_{\rm max} = 0.077 \times T_{\rm obs}^{1.19}
\end{equation}

The ratio between $\tau_{\rm LC}$, the observed $e$-folding time of
the flare's light curve (determined by fitting the light curve from
the peak of the flare down to the 10\% of peak level) and the
thermodynamic decay $e$-folding time $\tau_{\rm th}$ is linked to the
slope $\zeta$ in the $\log \sqrt{EM}$--$\log T$ diagram by

\begin{equation}
  \label{eq:zeta}
  {\tau_{\rm LC} \over \tau_{\rm th}} = F(\zeta) =
  {c_a}e^{-\zeta/\zeta_a} + q_a  
\end{equation}
where the constants are, for the ASCA SIS detector, $c_a = 61$,
$\zeta_a = 0.035$ and $q_a = 0.59$.  The formula for the loop
semi-length $L$ is therefore:

\begin{equation}
  \label{eq:lzeta}
  L=\frac{\tau_{\rm LC} \sqrt{T_{\rm max}}}{\alpha F(\zeta)}  ~~~~~~~~ 0.4 <
  \zeta \leq 1.7 
\end{equation}
where the second part of the relationship gives the range of $\zeta$
values allowed according to the modeling. The uncertainty on $L$ comes
both from the propagation of the errors on the observed parameters
$\tau_{\rm LC}$ and $\zeta$ and from the uncertainty intrinsic to the
modeling, i.e. the ability of Eq.~(\ref{eq:lzeta}) to reproduce the
true length of the modeled loops.  From the self-consistency checks
the latter error amounts to $\simeq 18$\%.

\begin{figure*}[!thbp]
  \begin{center}
    \leavevmode \epsfig{file=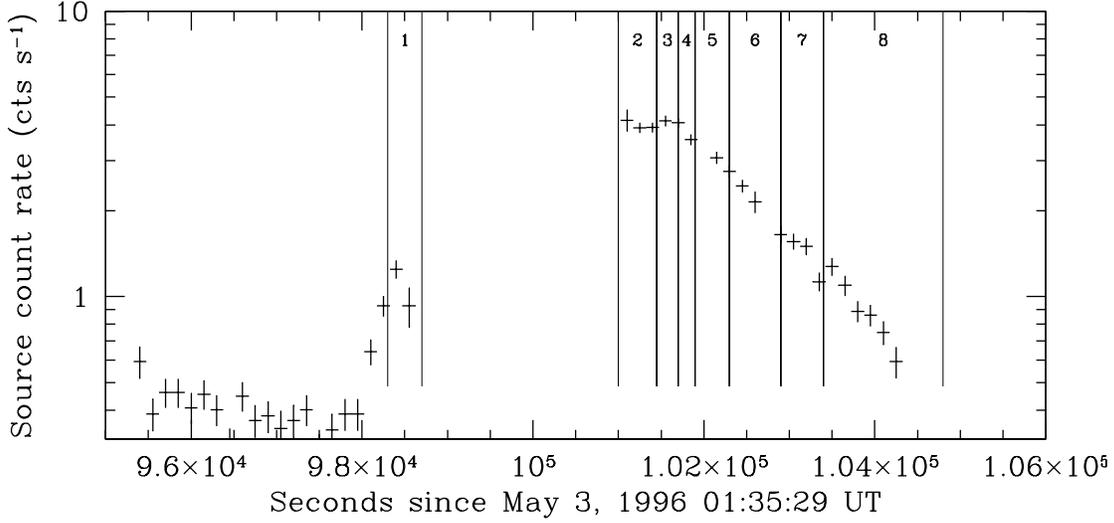, width=15.0cm, bbllx=20pt,
      bblly=400pt, bburx=600pt, bbury=700pt, clip=}
    \caption{The light curve of the second flare (``flare~2'')
      observed by the ASCA SIS on AD~Leo, background-subtracted
      (background count rate $\simeq 0.025$ cts s$^{-1}$) and binned
      in 150~s intervals. Also shown is the extent of the time
      intervals from which the spectra which have been used for the
      determination of the flare's spectral parameters have been
      extracted.}
    \label{fig:lcflare}
  \end{center}
\end{figure*}

\begin{figure}[thbp]
  \begin{center}
    \leavevmode \epsfig{file=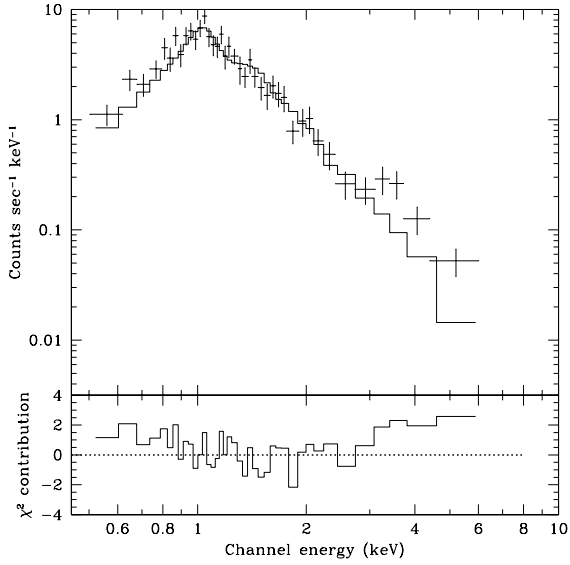, width=8cm, bbllx=25pt,
      bblly=150pt, bburx=600pt, bbury=700pt, clip=}
    \caption{The SIS-0 spectrum of AD Leo at the peak of flare~2 (time
      interval~2 in Fig.~\ref{fig:lcflare}, plotted together with the
      best-fit {\sc mekal} isothermal model fit to the flaring
      emission.}
    \label{fig:flare2_2}
  \end{center}
\end{figure}

We have applied the above approach to the three flares visible in the
ASCA light curve of AD~Leo; for each of them the light-curve has been
subdivided in time intervals containing $\simeq 1000$~cts per SIS
spectrum, as shown in Fig.~\ref{fig:lcflare} for the second event. The
SIS-0 and SIS-1 spectra from each of these events have been
simultaneously fit with a one-temperature {\sc mekal} model with
freely varying parameters, plus a two-temperature {\sc mekal} model
with parameters fixed at the best-fit values for the quiescent
emission. An example of the resulting fit is shown in
Fig.~\ref{fig:flare2_2}, for the spectrum at the peak of the SIS
flare~2. The abundance for the flaring component was fixed at the same
value as for the quiescent component (including the same level of
$\alpha$ element enhancement), and best-fit temperatures and
emission-measures have been derived. The limited statistics of the
individual flaring spectra do not allow for the metal abundance of the
flaring components to be separately determined. If it is left as a free
parameter the resulting confidence range is too wide to yield useful
information, and is in each case compatible with the quiescent
abundance.

\begin{table*}[htbp]
  \caption{The spectral parameters $T$ and $EM$ derived for the
    individual phases of the second flare in the AD~Leo ASCA
    observation from the analysis of the SIS spectra. The spectra have
    been analyzed with a single-temperature model (plus a
    frozen-parameter two-temperature model to account for the
    quiescent emission).  The bounds of the confidence intervals (at
    $\Delta \chi^2 = 2.7$) are also reported for each parameter. The
    time interval $i$ to which each set of parameters applies is
    shown in Fig.~\ref{fig:lcflare}.}
  \begin{center}
    \leavevmode
    \begin{tabular}{r|rrr|rrr|rrr}
      $i$ & $T$ & $T_{-90\%}$ & $T_{+90\%}$  &
      $EM$ & $EM_{-90\%}$ & $EM_{+90\%}$
      & $\chi^2$ & DoF & Prob.\\ 
      & \multicolumn{3}{c|}{keV}  & 
      \multicolumn{3}{c|}{$10^{50}$ cm$^{-3}$} & 
      & & \\\hline
1 & 2.12 & 1.48 & 3.44 &  84.8  &  68.3 & 100.6 & 0.95 & 25 & 0.53\\
2 & 1.27 & 1.19 & 1.36 &  509.9 & 484.4 & 535.6 & 1.24 & 71 & 0.08\\
3 & 1.33 & 1.24 & 1.43 &  501.5 & 476.8 & 526.1 & 1.32 & 78 & 0.03\\
4 & 1.31 & 1.21 & 1.43 &  471.6 & 444.9 & 498.7 & 1.15 & 61 & 0.20\\
5 & 1.10 & 1.03 & 1.18 &  361.1 & 339.6 & 382.8 & 1.23 & 59 & 0.11\\
6 & 1.09 & 1.02 & 1.17 &  276.1 & 260.3 & 292.0 & 1.07 & 65 & 0.33\\
7 & 1.00 & 0.91 & 1.09 &  139.0 & 128.5 & 149.6 & 1.28 & 55 & 0.08\\
8 & 0.78 & 0.71 & 0.85 &  70.89 &  64.3 &  77.5 & 0.83 & 61 & 0.83\\
    \end{tabular}
    \label{tab:flaring}
  \end{center}
\end{table*}

\subsubsection{ASCA flare~1}

For the first of the three ASCA flares too few source photons are
available to allow a complete analysis. With a peak count rate of $\la
2$~cts~s$^{-1}$ and a decay time scale $\tau_{\rm LC} \simeq 570$~s,
it is not possible to follow the decay of the event in the $\log
\sqrt{EM}$--$\log T$ plane. The peak temperature for the event is
$\simeq 20$~MK. An upper limit to the loop size can be obtained
through Eq.~(\ref{eq:lzeta}) by assuming that the influence of
sustained heating is negligible (i.e.\ $F(\zeta) \simeq 1$). Under
this assumption, $L < 10^{10}$~cm. The peak X-ray luminosity for the
event is $9 \times 10^{28}$~erg~s$^{-1}$.

\subsubsection{ASCA flare~2}

The second flaring event, the better defined of the three, has
sufficient source counts to allow a detailed analysis and to constrain
rather narrowly the characteristics of the flaring region. Its
light-curve is shown in Fig.~\ref{fig:lcflare}, where the intervals in
which the flare has been subdivided for the spectroscopic analysis are
shown.

\begin{figure}[!thbp]
  \begin{center}
    \leavevmode \epsfig{file=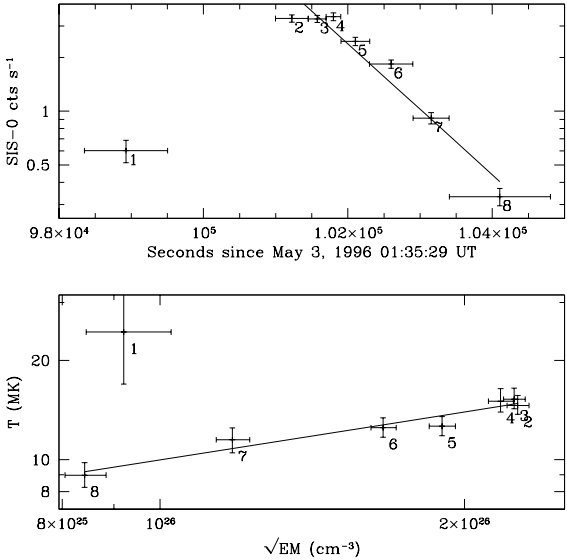, width=8cm, bbllx=25pt,
      bblly=150pt, bburx=600pt, bbury=700pt, clip=}
    \caption{The top-panel shows the temporal evolution of the
      light-curve of the second flare seen in the SIS light-curve
      (flare~2), binned in intervals containing $\simeq 1000$ SIS
      counts, together with the best-fit exponential decay to the
      light-curve (yielding an $e$-folding time of $1180 \pm 130$~s).
      The bottom panel shows the $T$ versus $\sqrt{EM}$ plot for the
      same time intervals. The flare decay slope in this plane is
      $\zeta = 0.48 \pm 0.06$.}
    \label{fig:flare2}
  \end{center}
\end{figure}

The evolution of the flare in the $\log \sqrt{EM}$--$\log T$ plane is
shown in the lower panel of Fig.~\ref{fig:flare2}, together with a
least-square fit to the decay phase. The resulting best-fit slope is
$\zeta = 0.48 \pm 0.06$. The top panel of Fig.~\ref{fig:flare2} shows
the light curve binned in the same intervals as used for the spectral
analysis, together with the best-fitting exponential decay (with an
$e$-folding time $\tau_{\rm LC} = 1180 \pm 130$~s).

Application of Eq.~(\ref{eq:zeta}) yields a ratio between the observed
cooling time scale $\tau_{\rm LC}$ and the thermodynamic cooling time
scale for the flaring loop $\tau_{\rm th}$ of $F(\zeta) = 5.4$,
indicative of the presence of strong sustained heating, and showing
that the observed decay is driven by the time-evolution of the heating
process and not by the spontaneous decay of the loop.  As $\tau_{\rm
  LC} = 1.18 \pm 0.13$~ks, $\tau_{\rm th} \simeq 220$~s.  The
intrinsic flare peak temperature is, applying Eq.~(\ref{eq:tobs}) to
the observed maximum temperature, $T_{\rm max} \simeq 48$~MK.  From
Eq.~(\ref{eq:lzeta}) the loop semi-length is then $L=(4.1 \pm 1.0)
\times 10^{9}$~cm, i.e.\ $L \simeq 0.15~R_*$. This loop length is much
smaller than the pressure scale height\footnote{defined as $H =
  2kT/\mu g \simeq 6000 \times T_{\rm max}/(g/g_\odot)$, where T is
  the plasma temperature in the loop, $\mu$ is the molecular weight of
  the plasma and $g$ is the surface gravity of the star.} $H\simeq 3
\times 10^{11} \simeq 11~R_*$.

The peak X-ray luminosity of the flare is $\simeq 2.5 \times 10^{29} ~
\rm erg ~ s^{-1}$ in either the 0.5--10~keV or the 0.5--4.5~keV band,
and the total energy radiated in X-rays by the flare (obtained by
simply integrating the X-ray luminosity in each of the intervals in
which the flare has been subdivided) is $\simeq 3.2 \times
10^{31}$~erg.  Using the same simple argument discussed by
\cite*{frm+99} we can estimate the magnitude of the magnetic field
necessary to confine the loop and to produce the energy (presumably by
magnetic recombination) radiated in the flare. The magnetic field
required to confine the flaring plasma (which has a maximum pressure
of $\simeq 2 \times 10^4$~dyne~cm$^{-2}$) is $B \simeq 0.6$~kG, while
the minimum magnetic field required to explain the flare energetics is
$B \simeq 1.4$~kG.  Such fields are relatively modest, and fully
compatible with the values of few kG with large filling factors
measured e.g. by \cite*{jv96} in a few active M dwarfs with
characteristics similar to AD~Leo.

\subsubsection{ASCA flare~3}

The decay of the third flare is more irregular than for flare~2, and
with a lower statistics (given that the peak count rate is a factor of
$\simeq 2$ lower), resulting in a much larger range for the slope $T$
versus $\sqrt{EM}$ plot, and thus in a much larger confidence region
for the estimate of the loop length. The nominal observed decay time
is $\tau_{\rm LC} \simeq 3500$~s, and the peak temperature is 38~MK.
Also in this case there is evidence for strong sustained heating (with
$\zeta \simeq 0.4$, at the limit of validity of the method, implying
$\tau_{\rm LC}/ \tau_{\rm th} \simeq 8$), and the resulting loop size
is compatible with the one derived for flare~2, at $\simeq 7 \times
10^9$~cm, with a $1\,\sigma$ uncertainty of a factor of $\simeq 2$.
The intrinsic decay time is, at $\tau_{\rm th} \simeq 400$~s, similar
to the one of flare~2. The compatible loop size, together with the
temporal proximity of the two events is in principle consistent with
the two (or even three) flares coming from the same loop, repeatedly
heated. However, ``sympathetic flaring'', as observed in the Sun is
also a possibility.

\subsection{The PSPC flares}
\label{sec:pspc}

The first PSPC flare (as reported by \cite{rm98}) has a short decay
time $\tau_{\rm LC} = 680$~s, with evidence for moderate sustained
heating ($F(\zeta) \simeq 2.5$), so that the loop intrinsic decay time
is $\tau_{\rm th} \simeq 300$~s, very similar to the decay time of the
better defined of the events analyzed here (the ASCA flare~2). The
loop length derived by \cite*{rm98} is $L \simeq 4 \times 10^9$~cm,
with $\simeq 60$\% uncertainty. The peak temperature is $\simeq
35$~MK.

For the purpose of the analysis the light-curve of the second flare
(which we analyze in the present paper) has been subdivided in 5
intervals, of duration 1, 1, 1, 1.2 and 1.7~ks.  The event is a
long-lasting one, with $\tau_{\rm LC} \simeq 8.6$~ks (i.e.\ the
longest among the events analyzed here); the slope of the flare decay
is in this case $\zeta_{\rm PC} = 1.8$, outside the domain of validity
of the method (see \cite{rm98}), and implies $\tau_{\rm H} \gg
\tau_{\rm lc}$, i.e. the decay is totally dominated by the heating
profile, while the effects of the loop decay are negligible. The
intrinsic characteristics of the flaring loop are thus ``hidden'' by
the time evolution of the heating, and cannot be effectively
constrained by studying the decay light-curve.  It is however still
possible to derive an upper limit to the loop's size: the uncertainty
on $\zeta_{\rm PC}$ is large (due to the irregular decay of the event,
with visible evidence for reheating events, during which the count
rate rises again during the decay phase), and its upper bound enters
among the allowed range of values for $\zeta_{\rm PC}$, allowing to
derive an upper limit, $L < 17 \times 10^9$~cm, still well below the
pressure scale height.

\subsection{The \emph{Einstein} IPC flare}
\label{sec:ipcflare}

We have analyzed this long-lasting event with the same approach as the
one described by \cite*{rm98} for flares observed with the PSPC
detector, but tuned for the IPC detector (the details are described in
\ref{app:ipctech}).  The light-curve of the flare (i.e.\ the second
and third observation segments visible in Fig.~\ref{fig:ipclc}) has
been divided, for the purpose of the analysis, in 4 intervals.
Segment~2 has been subdivided in three intervals (of 600, 800 and
1060~s each), while segment~3 was considered as a single interval of
1550~s duration. The length of these intervals was chosen so to have
approximately the same number of photons ($\simeq 4000$) in each
spectrum. The quiescent spectrum extracted from the 2700~s pre-flare
segment, properly scaled, has been subtracted from the spectra
extracted from the four flaring time intervals to derive the net flare
spectra.

The IPC event has an observed decay time $\tau_{\rm LC} = 5400 \pm
600$~s; our analysis again shows that significant heating is present
during the decay phase, with $F(\zeta) = 3.8$, so that the
thermodynamic loop decay time is $\tau_{\rm th} \simeq 1400$~s. The
peak temperature is relatively low ($T_{\rm max} = 12$~MK), and the
loop length is $L = (1.3 \pm 0.4) \times 10^{10}$~cm (or $0.5 \pm
0.15~R_*$).

\section{Discussion}
\label{sec:disc}

Under the assumption of validity of the method used to derive
parameters of the flaring loops (and discussed in detail in
\cite{rbp+97}), our comparative analysis of flares on AD~Leo (whose
results are summarized in Table~\ref{tab:flares}), with six events
observed by three different instruments, allows -- together with the
study of the quiescent emission as seen by the ASCA SIS -- for the
first time to study the characteristics of the AD~Leo corona (and by
extension of the coronae of flare stars) in a systematic way. 

\subsection{Physical parameters of the flaring loops}

All the flares we have studied point toward the structures in the
flaring component of the corona of AD~Leo being confined to a rather
narrow range of characteristic sizes, small in comparison with the
stellar radius ($L \la 0.5~R_*$). A summary of the characteristics of
each flare analyzed here is shown in Table~\ref{tab:flares}. No
evidence for large loops, extending out to distances of one or more
stellar radii is found in any of the observations studied. The peak
flare temperatures range between $T \simeq 10$~MK and $T \simeq
50$~MK, and the peak X-ray luminosity is comprised in a narrow range
($\simeq 1$--$3 \times 10^{29}~ \rm erg ~ s^{-1}$).  The total
radiated energy is $\simeq 1$--$25 \times 10^{32}$~erg, a much larger
range than the peak luminosity, due to the large range of decay times.
The pressures also span two orders of magnitude, from $\simeq 100$ to
$\simeq 10\,000$~dyne~cm$^{-2}$.

\begin{table*}[htbp]
    \caption{Characteristics of the various flaring events studied on
      AD~Leo. The stellar radius is $R_* = 26 \times 10^9$~cm. For
      each flare we list the observed light-curve decay time, the
      value of $F(\zeta)$ resulting from Eq.~(\ref{eq:zeta}), the
      corresponding intrinsic loop thermodynamic decay time, the loop
      semi-length inferred from Eq.~(\ref{eq:lzeta}), the peak
      temperature, the peak emission measure and X-ray luminosity (in
      the 0.5--4.5~keV band) and the total energy released in X-rays
      during the decay of the event, the pressure resulting from the
      application of the Rosner et~al.\ (1978) scaling laws
      for the loop length and temperature derived from the flare
      analysis, and the loop volume, density and pressure derived from
      the peak $EM$ and $T$ and assuming a loop aspect ratio $\beta =
      0.1$ (typical of solar loops). The subscript within parentheses
      indicates the power of 10 by which the given quantity has been
      scaled.}
  \begin{center}
    \leavevmode
    \begin{tabular}{rrrrrrrrrrrrr}
&&&&&&&&&&\multicolumn{3}{c}{$\beta = 0.1$}\\\cline{11-13}
Flare & $\tau_{\rm LC}$ & $F(\zeta)$ & $\tau_{\rm th}$ & $L_{(9)}$ & $T_{\rm
     max}$ & $EM_{(50)}$ & $L_{\rm X\,(28)}$ & $E_{\rm X\,(31)}$ &
     $P_{\rm RTV\,(2)}$ & $V_{(27)}$ & $n_{(11)}$ & $P_{(2)}$\\ 
    & s & & s & cm  & MK & cm$^{-3}$ & erg~s$^{-1}$ & erg &
     dyn~cm$^{-2}$ & cm$^{3}$ & cm$^{-3}$ &dyn~cm$^{-2}$ \\\hline 
 SIS--1 &  570 & $> 1$ & $<570$ &  $< 10.0$ & 20 & 140 & 9 & 13
 & $> 3$ & $< 60$ & $> 5$ & 13 \\
 SIS--2 & 1180 & 5.4 & 220 &  $4.1\pm 1.0$  & 48 & 510 & 25 & 82 &
 100 & 4.3 & 34 & 200 \\
 SIS--3 & 3500 & 8.0 & 400 &  $7.0\pm 3.5$  & 38 & 250 & 7 & 16 & 29
 & 22 & 11 & 56 \\
PSPC--1 &  680 & 2.5 & 300 &  $4.0\pm 2.4$  & 35 & 200 & 18 & 12 &
 40 &  4 & 20 & 200 \\
PSPC--2 & 8040 & $\gg 1$ &   -- &  $< 17.0$ & 12 &  80 & 19 & 25 &
 $>0.4$ & $\la 300$ & $>1.5$ & $>5$ \\
IPC     & 5400 & 3.8 & 1400 & $13.0\pm 4.0$ & 12 & 100 & 32 & 250 &
 0.5 & 140 & 3 & 10 \\ 
    \end{tabular}
    \label{tab:flares}
  \end{center}
\end{table*}

The only previously published analysis of flares on AD~Leo is the one
of \cite*{cfh+97}, who studied a significant (although smaller than
the event discussed here) EUVE flare, with simultaneous ground-based
optical coverage. Their analysis is however based on the assumption
that during the decay phase the heating rate is negligible in
comparison with the natural cooling rate of the flaring loop, so that
the light curve is fully dominated by the spontaneous cooling of the
loop (i.e.\ $\tau_{\rm LC} \simeq \tau_{\rm th}$, an approximation
which \cite{hfs+95} call the ``strong condensation limit'') and
consequently long rise and decay times require large loop lengths
(\cite{hfs+95}).  The EUVE spectrum alone (given the very low $S/N$ of
the continuum) does not allow to derive the coronal abundance, and
thus \cite*{cfh+97} derive the flare parameters under the assumption
of $Z = Z_\odot$ and $Z = 0.1 \times Z_\odot$ coronal metallicity.
The loop length they derive (for the first of the two flares they have
observed) is $L = 4.7$--$1.5 \times 10^{10}$~cm (the first value is
for $Z = Z_\odot$, the second for $Z = 0.1 \times Z_\odot$, as in the
following), the density $n = 4$--$40 \times 10^9$~cm$^{-3}$, the
pressure $P = 10$--$100$~dyne~cm$^{-2}$, with an equipartition
magnetic field $B = 16$--$50$~G. The peak temperature is (independent
of the assumed abundance) $T = 13$~MK.

The resulting loop lengths are thus significantly larger (4 to 11
times) than the lengths derived from the ASCA flare~2 (the one which
gives the tightest constraint to the derived physical parameters).
Although these are distinct events, and thus could in principle result
from quite different types of flaring structures, it is worth nothing
that for all the X-ray flares analyzed in the present paper there is
evidence for substantial heating during the decay phase (with $
\tau_{\rm LC}/ \tau_{\rm th}$ ranging from $\simeq 2.5$ to $\simeq
8$); if these flaring events had been analyzed assuming that
$\tau_{\rm LC} \simeq \tau_{\rm th}$ the resulting loop lengths would
have been over-estimated by factors 2.5--8; on the assumption that the
EUVE flare is of a similar class as the ones observed in X-ray it is
likely that equating $\tau_{\rm th}$ to $\tau_{\rm LC}$ will lead to
an over-estimate of the loop length by similar factors as in the X-ray
case. The large ($L \simeq R_*$) loop sizes derived by \cite*{cfh+97}
for the EUVE flares would in this case be reduced to the same
characteristic size ($L \simeq 0.3 \times R_*$) derived for the X-ray
flares, thus again pointing to a narrowly confined characteristic size
for the structures of the flaring component of the AD~Leo corona.

\subsection{The aspect ratio of the flaring loops}

Under the assumption that the flaring loops are similar to solar loops
(i.e.\ that their physical parameters are regulated by the same
scaling laws of \cite{rtv78}) we can derive an estimate for the ratio
for the loop's cross-section radius and its length (which in the solar
case is typically $\beta \simeq 0.1$, although with a wide range of
observed values). This can be done by comparing the pressure derived
through the scaling laws with the pressure derived from the flare's
emission measure. In the former case the pressure is obtained by
application of the \cite*{rtv78} scaling law

\begin{equation}
\label{eq:rtv}
T_{\rm max} = 1.4 \times 10^3 (P_{\rm RTV} L)^{1/3}
\end{equation}
(which can also be written as $T_{\rm max} = 6.16 \times 10^{-4}
\sqrt{n_{\rm RTV} L}$) where $P_{\rm RTV}$ is the pressure at the base
of the loop, and where the flare's peak temperature and emission
measure are used.  In the latter case a simple estimate of the density
of the flaring plasma can be obtained by dividing the peak emission
measure by the volume of the flaring loop, where the volume is
obtained by assuming a given value for $\beta$, so that

\begin{equation}
  \label{eq:dens}
  n = \sqrt{{ {EM} } \over {2 \pi L^3 \beta^2}}
\end{equation}
and

\begin{equation}
  \label{eq:press}
  P = nkT = kT \times \sqrt{{ {EM} } \over {2 \pi L^3 \beta^2}}
\end{equation}

Under the assumption that the flaring loop at flare maximum is not far
from a steady-state condition (i.e.\ that it is almost filled up with
flaring plasma), the two estimates for the pressure should yield
similar results. In Table~\ref{tab:flares} we report the resulting
estimates for each individual flare discussed here, with the pressure
and density estimates obtained under the assumption that the loops are
similar to the ones observed in the solar-corona, where typically
$\beta \simeq 0.1$. As it can be seen the assumption of solar-like
aspect ratio loops yields pressures which are systematically larger
(and thus volumes that are systematically smaller) than the pressures
derived from the scaling laws. The factors range from $\simeq 2$ (for
the ASCA-2 flare) up to $\simeq 20$ (for the IPC flare). The two can
be reconciled under the assumption of higher $\beta$ values, implying
loops which are ``fatter'', with a resulting $\beta \simeq 0.3$.

The larger $\beta$, if generally applicable, implies lower densities,
so that even at the peak of the largest observed flares the plasma
likely remains optically thin, justifying the analysis of the spectra
without any allowance for self-absorption. In the case of the large
EV~Lac flare discussed by \cite*{frm+99} the larger $\beta$ would for
example imply that the average peak density is $n \la
10^{12}$~cm$^{-3}$.

\subsection{Energetics}

A simple estimate of the heating released during the flare decay can
be obtained as follows: if one assumes uniform heating along the loop
the scaling laws of \cite*{rtv78} yield

\begin{equation}
  \label{eq:rtvh}
  \frac{d H}{d V d t} \simeq 10^5 ~ P_{\rm RTV}^{7/6} ~ L^{-5/6}
\end{equation}
and the total heating rate at the flare maximum is therefore
\begin{equation}
  \frac{d H}{d t} \simeq \frac{d H}{d V d t} \times V 
\end{equation}

Given that all the events analyzed here display significant heating
during the decay, the heating time scale can be approximated as
$\tau_{\rm H} \simeq F(\zeta) \times \tau_{\rm th} \simeq \tau_{\rm
  LC}$, and an approximate estimate of the total energy released
during the flare decay is 

\begin{equation}
  E \simeq \frac{d H}{d t} \times \tau_{\rm LC}.
\end{equation}

If the above analysis is applied to the the flares we have studied
(determining the volume on the assumption of $\beta \simeq 3$) the
peak volumetric heating rates range (for the flares for which the
length is determined) between $\simeq 5$ and $\simeq 50~ \rm erg ~
cm^{-3} ~ s^{-1}$, and both the peak total heating and the total
energy released during the flare are a few times ($\simeq 3$) larger
than the peak X-ray luminosity and total radiated energy, showing that
radiative losses are not the dominant term in the flare energetics
(similarly to what determined for the large ASCA flare on EV~Lac,
\cite{frm+99}).

\subsection{The quiescent active corona of AD~Leo}

In the Sun the same general class of structures which constitute the
active part of the corona produce the flares. On the assumption that
this is also true in the stellar case, we can use the range of derived
properties for the flaring loops of AD~Leo to infer the
characteristics of the quiescent active corona, and in particular to
understand the scaling from the solar picture toward much higher
activity levels, i.e.\ whether this happens (mainly) through an
increase in the filling factor, in the characteristic size of the
coronal structures or through an increase in the pressure of the
plasma filling the loops. The simple picture of a straightforward
increase in filling factor, until the star is completely covered with
solar-like active regions, can rather naturally explain (as discussed
in detail by \cite{dpo+99}) the luminosity of ``intermediate
activity'' stars: using Yohkoh solar images \cite*{dpo+99} show that
the X-ray luminosity of the solar active regions is $\simeq 6 \times
10^{26}~\rm erg~s^{-1}$, with a filling factor $f \simeq 1$--2\%.
Thus, by covering a solar-type star with active regions a luminosity
of $\simeq 3$--$6 \times 10^{28}~\rm erg~s^{-1}$ can be achieved.
However, such scheme, while appealing in its simplicity, is in
contrast with some of the available spectroscopic indicators of
coronal structuring; \cite*{vmp98} for example have analyzed the ROSAT
PSPC spectra of active solar-type stars, showing that they are
compatible with an active corona being constituted by compact,
high-pressure loops with filling factors of few percent (although
low-pressure, larger-pressure loops are also a possible solution).
Direct pressure measurements of coronal pressure have been performed
on a few stars with EUVE: while low- and intermediate-activity
solar-type stars show solar-like densities (e.g.\ $\epsilon$~Eri,
$n\simeq 10^9$ to $10^{10}$~cm$^{-3}$, \cite{sds+96}; Procyon,
$n\simeq 3 \times 10^9$~cm$^{-3}$, \cite{sdh+96}), supporting
\cite*{dpo+99}'s view of an increased filling factor being responsible
for higher densities, more active stars (active binaries) show
evidence for much increased coronal density ($n \ga 10^{13}$~cm$^{-3}$
e.g. in 44~Boo, \cite{bd98}), well above the value required to explain
the enhanced X-ray luminosity through an $f \simeq 1$ corona, and
again implying a small $f$ in the more active stars.

For a dMe star such as AD~Leo, with a smaller surface area (by a
factor of 7 with respect to the Sun) the maximum quiescent X-ray
luminosity, if covered with solar-type active regions is $\simeq
(4$--$8) \times 10^{27}~\rm erg~s^{-1}$, i.e.\ still an order of
magnitude below the observed values. Thus, another mechanism needs to
be postulated to explain the higher X-ray luminosity.  Given the
compact sizes derived here for the flaring regions (comparable in
relative terms to the solar ones), there is no evidence for large
loops with significant emissivity and thus for a corona extending to
large distances from the star and occupying a large volume. Given the
low filling factor we derive for the quiescent component (see below)
the additional X-ray emissivity is therefore likely to come from an
increase in plasma pressure within the same types of (relatively
small) coronal structures as in solar active regions. This is also
compatible with the high photospheric magnetic fields (few kG)
measured in dMe stars (\cite{jv96}), which can easily confine the
higher pressure loops required by this picture of the corona.

\cite*{grk+96} have studied the ROSAT spectra and light curves of
several M dwarfs (including AD~Leo) with the help of semiempirical
loop models. They infer that two distinct thermal components are
present in the coronae of M dwarfs, a quiescent cooler one showing
negligible time variability, and a hotter flaring one responsible for
the observed variability. The cooler component has a characteristic
temperature (in the PSPC spectra) of $\simeq 3$~MK, while the hotter
one has $T \simeq 10$~MK. The cool component can be modeled in terms
of small ($L \ll R_*$), high-pressure ($P \ga P_\odot$) loops, while
for the hotter component two classes of solutions are possible, i.e.\ 
it can be composed either of large ($L \ga R_*$), high-pressure ($P
\ga P_\odot$) loops or of small ($L \la R_*$) loops with a very high
pressure ($P \gg P_\odot$) and a very small filling factor ($f \ll
0.1$).  Again, the former possibility (a flaring component of the
corona composed of large loops) is not supported by our analysis; this
is also in agreement with the results of \cite*{smf+99}, who have
modeled the time-averaged SAX LECS and MECS spectrum of AD~Leo using
static loop models; they show that the observed spectrum is compatible
with being emitted from compact loops ($L \la 0.1~R_*$) with a small
filling factor ($f \la 10^{-3}$).

In the present case the filling factor for the hotter component can be
estimated under the assumption that the corona is largerly composed by
loops which satisfy the scaling laws of \cite*{rtv78} -- i.e.\ 
Eq.~(\ref{eq:rtv}) -- starting from the measured quantities $T$, $EM$
and $L$.  Assuming that the loop size (parameterized as $L = \alpha
\times R_*$, where $\alpha \simeq 0.3$) is now known, the emission
measure is linked to the loop size and density by

\begin{equation}
  \label{eq:em}
  EM = Nn^2V = N n^2 2 \pi \beta^2 (\alpha R)^3
\end{equation}
where $N$ is the number of emitting loops, $V = \pi \beta^2 \alpha^3
R^3$ is the volume of a single loop and $\beta$, as before, the
aspect ratio of the loop. If the filling factor is defined as the
fraction of the stellar area covered by loops, i.e.\ 

\begin{equation}
  \label{eq:f1}
  f = {{N A} \over {4 \pi R^2}}
\end{equation}
where $A = \pi \beta^2 \alpha^3 R^2$ is cross-section of an individual
loop, then Eq.~(\ref{eq:rtv}) can be used to express $f$ as a function
of the known quantities, i.e.\ 

\begin{equation}
  \label{f2}
  f \simeq 2 \times {{1.44 \times 10^{-13} \alpha EM} \over {4 \pi R
      T^4}}
\end{equation}
where the extra factor of 2 accounts for the fact that we only see one
hemisphere of the star at any given time.

Thus for the hot component of the quiescent emission seen in the ASCA
observation ($EM = 2.3 \times 10^{51}$~cm$^{-3}$), and assuming that
quiescent loops have the same size as flaring ones ($\alpha \simeq
0.3$) one can derive $f \simeq 6$\% (independent of $\beta$).  Simple
application of the scaling law (Eq.~(\ref{eq:rtv})) also yields
estimates for the density ($n \simeq 5 \times 10^{10}$~cm$^{-3}$) and
pressure ($P \simeq 70$~dyne~cm$^{-2}$) of the ``quiescent'' loops.
Our results therefore point to a corona in which the ``hot'' component
is composed of loops with comparable relative sizes as the solar
active region loops (i.e.\ $L \simeq 0.3~R_*$), but with significantly
higher pressure (by approximately an order of magnitude, typical solar
active region loop pressures being $P_\odot \simeq 5$~dyne~cm$^{-2}$).
The required filling factor is small, i.e.\ only a small fraction of
the stellar surface needs to be covered by such loops to explain the
observed emissivity.

\subsection{Metal abundance of the quiescent corona}

The initial results from the ASCA observations of coronal sources,
e.g.\ that the metal abundance of the coronal plasma is apparently
``non-solar'' has prompted a lively debate about the possible presence
of fractionation mechanisms which would deplete the plasma along its
way from the photosphere and chromosphere toward the corona.
Mechanisms of this type are apparently at work in the Sun, were
coronal abundance ratios are different from photospheric ones, with
elements selectively enhanced on the basis of their first ionization
potential (FIP). The solar corona however appears to be in general
terms to be metal-enriched, rather than metal-depleted. While there is
strong evidence, for example, that the coronal metal abundance changes
significantly during large flares (e.g.\ on Algol, \cite{sut+92};
\cite{os96}; \cite{fs99}), detailed assessments of the relative
coronal versus photospheric abundances in stellar coronae have been
hindered by the lack of detailed photospheric abundance analyses in
most active stars (even on Algol it's unclear if the quiescent coronal
abundance is higher or lower than the photospheric value). The
situation is further complicated by the often contradictory situation
when such analyses are available (e.g.\ the contrasting results about
photospheric abundances of active binaries, \cite{rgp93} versus
\cite{opg98}).

In the case of AD~Leo, the availability of a recent, high-dispersion
spectral analysis of the photospheric Fe abundance allows a detailed
comparison with the coronal abundance as derived from the X-ray
observation. The photospheric metallicity of AD~Leo (${\rm [Fe/H]} =
-0.75$, \cite{jla+96}) is at the low end of the disk population
abundance range.  The ASCA observation shows no evidence for
differences between the coronal and photospheric Fe abundance. The
coronal abundance ratios are clearly different from the solar
photospheric values (${\rm [Si/Fe]} \simeq {\rm [O/Fe]} \simeq 4$);
however, while neither Si nor O abundance have been determined in the
photosphere, ${\rm [\alpha/Fe]}$ (the abundance of elements such as Si
and O, the so-called $\alpha$ elements) is known to be generally
enhanced in low-metallicity stars, with ${\rm [\alpha/Fe]} \simeq 0.4$
at ${\rm [Fe/H]} \simeq -1.0$ (see the review of \cite{mcw97}), so
that even the ``anomalous'' abundance patterns of the AD~Leo corona
are in full agreement with the expected photospheric abundance ratios,
and therefore no chemical fractionation mechanism is required to
explain the observed abundances in the quiescent corona of AD~Leo.

\section{Conclusions}
\label{sec:concl}

Our results shows that the frequent flares which have been observed on
the ``typical'' dMe star AD~Leo have all taken place in similar,
compact coronal loops. In all cases sustained heating is present, and
the flare decay is dominated by the time evolution of the heating. As
a consequence, estimates of the size of the flaring regions found
assuming that the loop is decaying undisturbed will significantly
overstimate the loop size, by factors (for the AD~Leo flares) of
$\simeq 5$.

The number of observations is such to allow to estimate the frequency
of significant flares in AD~Leo: adding to the \emph{Einstein}, ROSAT
and ASCA observations discussed here with the EXOSAT observations
(which covered $\simeq 110$~ks\footnote{The EXOSAT high orbit resulted
  in continuous time coverage, unlikely the other low-Earth orbit
  satellites, which have a $\simeq 50$\% duty cycle, and thus the two
  data sets are not fully homogeneous, and the resulting flaring rate
  is only approximate.} and in which one flare was observed,
\cite{pts90}) and with the SAX observation ($\simeq 40$~ks exposure,
three flares, \cite{smf+99}), a total of 10 flares have been detected
on AD~Leo in a total of 480~ks, yielding an average flaring rate of
one flare every approximately 50~ks (which does not include the
several minor flaring events visibile both in the PSPC and in the SIS
light-curves).

The small range of sizes for the flaring loops (on the assumption that
the same class of loops is also the main component of the active
corona) implies a corona which scales to larger luminosities by
filling relatively compact loop structures (similar to the solar ones)
with progressively higher pressure plasma.  A pressure about 10 times
the typical value for solar active regions is sufficient, with a small
filling factor $f \simeq 6$\%. Such a scenario is well in agreement
with the compact high-pressure loops with $f \ll 1$ which both
\cite*{grk+96} and \cite*{smf+99} derive for AD~Leo as well as other
flare stars through the spectral analysis of the quiescent coronal
emission.

These conclusions are somewhat difficult to reconcile with the
scenario discussed by \cite*{dpo+99}, in which the intermediate
activity levels are reached by increasing the filling factor of
solar-like (in size and pressure) active regions up to $f \simeq 1$.
The approximate filling factor we derive is $f \ll 1$, and the large
emissivity is fully explained by the increase in pressure. The flare
emission can be explained through a significant (about two order of
magnitudes over the quiescent value) increase in pressure in the same
loops responsible for the quiescent emission. However \cite*{dpo+99}
based their conclusions on the observation of solar-type stars, where
in principle the scaling mechanism to high X-ray luminosity could be
different from the one in flare stars.

The discrepancy between the pressure derived for the flaring loops
through the application of the \cite*{rtv78} scaling laws and the
pressure determined from the peak emission measure and temperature
assuming loops with a solar-like aspect ratio ($\beta \simeq 0.1$)
points at the stellar loops having larger volumes for a given length
than in the solar case, which can be obtained with a larger $\beta$,
i.e.\ with ``fatter'' loops. A value of $\beta \simeq 0.3$ is
sufficient to reconcile the two estimates of pressure.

The upcoming high spectral resolution observations of the X-ray
emission from both the quiescent and flaring emission from flare
stars, to be performed with the upcoming \emph{Chandra} and XMM
observatories will allow direct measurements of the plasma pressure
for a range of temperatures using selected spectral diagnostics, and
thus to test the correctness of the framework presented in the present
paper.  Although high-resolution spectroscopy will certainly be a more
powerful method for the study of the coronal structuring of active
stars than the study of flare decay, the approach used here will still
be important in the \emph{Chandra}-XMM era as it will be applicable to
a much larger number of stars (hence spanning a broad range of stellar
parameters) than high-resolution spectroscopy. Also, the pressures
derived through high-resolution spectroscopy will be weighted by the
plasma emissivity and thus, if significant filamentation within loops
is present, can in principle be much higher than the average pressure
implied by the loop's length and emission measure; such evidence has
been reported in the solar case (e.g.\ \cite{fbm+96}).  Therefore,
comparison of the spectroscopically measured pressures with the
flare-derived emission-measures and lengths will constitute a valuable
diagnostic of structuring within a single coronal loop.

The quiescent X-ray luminosity of AD~Leo is remarkably constant across
almost 20~yr, i.e.\ between the \emph{Einstein} IPC observation in
1980 and the 1997 SAX observation it ranges from $\simeq 3 \times
10^{28}$~erg~s$^{-1}$ (in 1980) and $\simeq 5 \times
10^{28}$~erg~s$^{-1}$ (in 1996), a remarkably small range also given
the diversity of the various instruments used in the comparison. As
recently reviewed by \cite*{ste98b}, active stars show no evidence of
a cyclic behavior in their activity level, and indeed AD~Leo is no
exception.

\begin{acknowledgements}
 
  G.~Micela and F.~Reale acknowledge the partial support of ASI and
  MURST. This research has made use of data obtained through the High
  Energy Astrophysics Science Archive Research Center (HEASARC) Online
  Service, provided by the NASA/Goddard Space Flight Center.

\end{acknowledgements}


\appendix

\section{Physical characteristics of AD~Leo and other X-ray observations}
\label{app:targ}

AD~Leo (Gl~388) is classified as dM3e (\cite{hks94}), and its distance
(from the ground-based parallax, \cite{gj91} -- AD~Leo wasn't a
Hipparcos target) is 4.9~pc. No companions to it are known
(\cite{rg97}).  It is one of the few M dwarfs for which a high
resolution abundance analysis of its photospheric spectrum is
available: using state of the art model atmospheres and near-IR
high-resolution spectra \cite*{jla+96} have estimated the photospheric
parameters at $T_{\rm eff} = 3350$, [Z/H]$~= -0.75 \pm 0.25$ and $\log
g = 4.5$. Its measured rotational velocity ($v \sin i = 6.2 \pm
0.8$~km~s$^{-1}$) places AD~Leo among the tail of rare, fast rotating
M dwarfs (\cite{dfp+98}); the photometrically-determined rotational
period is 2.7~d (\cite{sh86}.

The bolometric luminosity of AD~Leo (\cite{dfp+98}) is $M_{\rm bol} =
8.85$ (corresponding to a bolometric luminosity $L_{\rm bol} = 8.7
\times 10^{31}$~erg~s$^{-1}$). The K-band absolute luminosity is $M_K
= 6.26$ (from the average apparent K-band magnitude reported by
\cite{leg92}, and using $d = 4.9$~pc), which, through the
mass-luminosity relationship of \cite*{bca+98}, yields $M =
0.40~M_\odot$. Using the models of \cite*{cb97}, the radius at $M =
0.40~M_\odot$ is $2.64 \times 10^{10}$~cm at ${\rm [M/H]} = 0$, and
$2.57 \times 10^{10}$~cm at ${\rm [M/H]} = -1$. Based on the
above-mentioned abundance determination we have assumed $R = 2.6
\times 10^{10}$~cm ($\simeq 0.37~R_\odot$). At this mass
(\cite{cb97}), stars are still expected to have a substantial
radiative core, so that the interior structure is still
``solar-like''.

\subsection{Other X-ray observations}
\label{app:prex}

The first X-ray flare on AD~Leo was seen with the HEAO~1 A-2 soft
X-ray detector (\cite{klm+79}; who also show that the identification
of AD~Leo with a quiescent HEAO~1 source by \cite{alg+79} was not
correct). The approximate flaring X-ray luminosity reported is $\simeq
10^{30}$~erg~s$^{-1}$, with a peak emission measure of $\simeq
10^{53}$~cm$^{-3}$. 

The \emph{Einstein} observation is discussed in
Sect.~\ref{sec:einstein}, while EXOSAT observed AD~Leo in four
different occasions (\cite{pts90}), either as a pointed target or
serendipitously, for a total of $\simeq 110$~ks. The quiescent X-ray
luminosity reported was $5 \times 10^{28}$~erg~s$^{-1}$, and one flare
was observed, with a rise time ($1/e$) of 10~min and a decay time of
60~min. The flare represented a factor of 12 enhancement over the
quiescent flux, with peak X-ray luminosity $L_{\rm X} = 4.2 \times
10^{29}$~erg~s$^{-1}$ and a total energy release estimated at $1.5
\times 10^{33}$~erg.

ROSAT observed AD~Leo with the PSPC detector on May 8, 1991. The
resulting average spectrum was analyzed by \cite*{grk+96} -- who used
the \cite*{rs77} plasma emission code -- and later re-analyzed by
\cite*{smf+99} using the {\sc mekal} plasma emission code. The average
PSPC spectrum (including the two flaring events) analyzed with the
{\sc mekal} code is compatible with a two-temperature model with
metallicity $Z = 0.1~Z_\odot$, best-fit temperatures of 6.6 and 16~MK,
and emission measures of $16 \times 10^{51}$~cm$^{-3}$ and $2.7 \times
10^{51}$~cm$^{-3}$, respectively.  The average (including flares)
X-ray luminosity is $L_{\rm X} = 9 \times 10^{28}$~erg~s$^{-1}$.  To
fit the spectrum a larger interstellar column density than compatible
with the EUV observations is required ($N({\rm H}) = 1.7 \times
10^{19}$~cm$^{-2}$), similarly to many other coronal sources.

AD~Leo was also the target of a day-long SAX observation on April 23,
1997, discussed by \cite*{smf+99}. Continuous variability is evident
in the light curve, with at least three recognizable individual
flaring events with enhancements over the ``quiescent'' count rate of
$\simeq 5$ times (in the 1.5--7.0~keV band) and decay times of $\simeq
0.8$, 4.0 and 4.4~ks, with insufficient counts for a detailed
analysis. The total spectrum (i.e.\ including the flaring events)
needs three thermal components to be well fit, with the first two
components at temperatures of 3.7 and 12~MK, and a third (effectively
unbound) component at $T \ga 100$~MK.  The average X-ray luminosity in
the 0.1--7.0~keV band is $4 \times 10^{28}$~erg~s$^{-1}$. The SAX
spectrum can also be fit with static loop models, resulting in two
classes of loops, one with $T_{\rm max} = 13$~MK and the second with
$T_{\rm max} \ga 100$~MK. The cooler loops are compact ($L \la
0.1~R_*$) and with a small filling factor ($f \la 10^{-3}$), while the
size and filling factor of the hotter loops in not constrained by the
fit.

Finally, a long EUVE observation of AD~Leo was performed in March
1993, during which two flaring events where seen, which have been
analyzed by \cite*{cfh+97}. Their analysis is discussed in detail in
Sect.~\ref{sec:disc}.

\section{Derivation of flare parameters for the \emph{Einstein} IPC}
\label{app:ipctech}

Although less sensitive, the \emph{Einstein} IPC detector is in many
respects similar to the ROSAT PSPC. The main difference is the lower
spectral resolution, with the incoming photon energy being coded in
only 15 energy channels. The approach taken for the analysis of flares
seen in IPC data closely mirrors the one used for the analysis of PSPC
data, described in detail in \cite*{rm98}. Here we will only describe
the IPC-specific differences, and refer the reader to \cite*{rm98} for
a detailed description.

The essential difference between the approach used for the ASCA SIS
data (for which spectral temperatures are derived from a fit to
isothermal models) and the one used for the PSPC and for the IPC is
the definition, for the latter two detectors, of a temperature
indicator based on a linear combination of the photon counts in each
detector channel. This ``spectral-shape index'' (SSI) is obtained
through a principal-component analysis, and is, for low-resolution
detectors, more robust than a fit to the spectrum. Similarly, the
square root of the count rate (CR2) is used as a proxy to the density,
so that the $\log n$--$\log T$ diagram (e.g.\ Fig.~\ref{fig:flare2})
is replaced by a ``CR2--SSI'' diagram.

In the case of the IPC the spectral-shape index has been obtained
taking the first 10 PI channels, of which the first 9 are considered
as the independent ones.  The values of the $<C_{i}>$ and
$\sigma_{C_{i}}$ coefficients and the weights to calculate the
spectral-shape index for the IPC (using Eqs.~(3) and~(4) of
\cite{rm98}) are listed in Table~\ref{tab:scaleipc}. The relationship
linking the ratio of the observed decay time $\tau_{\rm LC}$ and the
intrinsic thermodynamic decay time $\tau_{\rm th}$ with the slope
$\zeta$ determined in this case in the CR2--SSI diagram (the
equivalent of Eq.~(\ref{eq:lzeta})) is

\begin{equation}
  \tau_{\rm LC}/\tau_{\rm th} = c_{I} e^{(\zeta/\zeta_{I})} + q_{I} =
  F_{I} (\zeta)
\label{eq:ipcssi}
\end{equation}
with:

\[
c_{I} = 5.91  ~~~ \zeta_{I} = -0.34 ~~~ q_{I} = 1.74
\]

For large flaring loops (i.e.\ loops larger than the local pressure
scale height) the relationship is:

\begin{equation}
  \tau_{\rm LC}/\tau_{\rm th} = c'_{I} e^{(\zeta/\zeta'_{I})} + q'_{I}
  = F'_{I} (\zeta)
\label{eq:ipcssi2}
\end{equation}
with
\[
c'_{I} = 76.9 ~~~ \zeta'_{I} = -1.30 ~~~ q'_{I} = 1.52
\]

\begin{figure}
\centerline{\epsfig{file=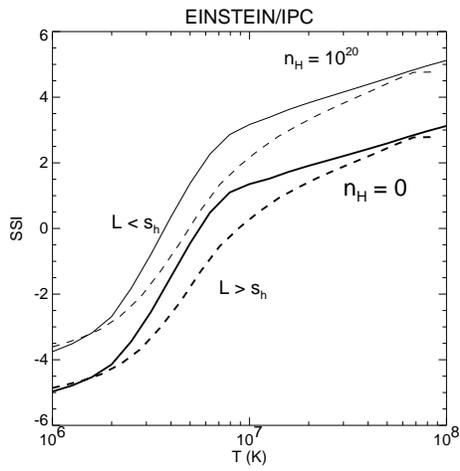,width=8.8cm}}
\caption[]{The Spectral Shape Index (SSI) of \emph{Einstein} IPC
  spectra as a function of the maximum temperature of hydrostatic loop
  models (cf. Fig.~1 of \cite{rm98}) with length much shorter ({\it
    solid lines}) and longer ({\it dashed lines}) than the local
  pressure scale height ($s_h$).}
\label{fig:ipcssit}
\end{figure}

Fig.~\ref{fig:ipcssit} shows the dependence of the IPC SSI on the
temperature at the top of model hydrostatic loops, analogous to Fig.~1
of \cite*{rm98}, to be applied to infer the flare maximum temperature,
which is to be used in Eq.~(\ref{eq:lzeta}) for deriving the loop
length.

The distribution of predicted vs. true model loop lengths is well
centered around the correct value (median $+~2$\%) with a standard
deviation of 18\%. Simulations analogous to those described in
Sect.~2.4 of \cite*{rm98} show that an uncertainty between 10\% and
30\% (for 10\,000 cts per bin) must be added, depending on the heating
time scale $\tau_{\rm H}$ (going from $\tau_{\rm H} = 0$ to $\tau_{\rm
  H} = 2 \tau_{\rm th}$). For 1000 cts per bin the uncertainties are
much larger than with PSPC ($> 80\%$), likely because the lower
spectral resolution induces a large indetermination in the slope in
the CR2--SSI diagram.

\begin{table}
  \caption{The coefficients (mean values, standard deviations and
    weights) for the computation of the \emph{Einstein} IPC spectral
    shape index using Eqs.~(3) and~(4) of Reale \& Micela (1998). The
    usage of PI channels is assumed.}
  \label{tab:scaleipc}
  \begin{center}
    {\footnotesize
      \begin{tabular}
        {c c c r | c c c r}
        C & $<C>$ & $\sigma_{C}$ & $W$ & C & $<C>$ & $\sigma_{C}$ & $W$\\
        \hline
        1&.1646 & .0634 &  0.36 &6&.0825 & .0301 & $-0.36$ \\
        2&.1931 & .0548 &  0.36 &7&.0559 & .0272 & $-0.36$ \\
        3&.1731 & .0228 &  0.33 &8&.0360 & .0224 & $-0.35$ \\
        4&.1456 & .0178 & $-0.18$ &9&.0222 & .0169 & $-0.33$ \\
        5&.1139 & .0280 & $-0.33$ & & & & \\
      \end{tabular}
      }
  \end{center}
\end{table}

\end{document}